\documentclass[10pt,onecolumn]{IEEEtran}
\usepackage{amsmath,amsthm,amssymb}
\usepackage{enumerate}
\usepackage{color}
\usepackage{url}
\usepackage{balance}
\usepackage{graphicx} 
\usepackage{mathrsfs}
\usepackage{epsfig}
\usepackage{verbatim}
\usepackage{setspace}
\usepackage{stfloats}
\newtheorem{theorem}{Theorem}
\newtheorem{lemma}[theorem]{Lemma}

\newtheorem{claim}[theorem]{Claim}
\newtheorem{remark}[theorem]{Remark}
\renewcommand{\vec}[1]{\mathbf{#1}}

\newcommand{\bbP}{\mathbb{P}}
\newcommand{\bbE}{\mathbb{E}}
\newcommand{\bbD}{\mathbb{D}}

\newcommand{\cT}{\mathcal{T}}
\newcommand{\cP}{\mathcal{P}}
\newcommand{\cX}{\mathcal{X}}
\newcommand{\ctdX}{\wtd{\mathcal{X}}}
\newcommand{\cZ}{\mathcal{Z}}
\newcommand{\cU}{\mathcal{U}}

\newcommand{\cS}{\mathcal{S}}
\newcommand{\cJ}{\mathcal{J}}
\newcommand{\cY}{\mathcal{Y}}
\newcommand{\cB}{\mathcal{B}}
\newcommand{\cC}{\mathcal{C}}

\newcommand{\cL}{\mathcal{L}}

\newcommand{\bX}{\mathbf{X}}
\newcommand{\bZ}{\mathbf{Z}}
\newcommand{\bY}{\mathbf{Y}}
\newcommand{\bJ}{\mathbf{J}}
\newcommand{\bU}{\mathbf{U}}
\newcommand{\btdX}{\mathbf{\wtd{X}}}

\newcommand{\bx}{\mathbf{x}}
\newcommand{\bs}{\mathbf{s}}

\newcommand{\bT}{\mathbf{T}}
\newcommand{\bz}{\mathbf{z}}
\newcommand{\by}{\mathbf{y}}
\newcommand{\bj}{\mathbf{j}}
\newcommand{\bu}{\mathbf{u}}

\newcommand{\td}[1]{\tilde{#1}}
\newcommand{\wtd}[1]{\widetilde{#1}}

\newcommand{\ajblue}[1]{\textcolor{black}{#1}}

\newcommand{\removed}[1]{}

\newcommand{\nb}{(n)} 

\newcommand{\sT}{\mathscr{T}}

\newcommand{\sQ}{\mathscr{Q}}

\newcommand{\done}{\delta_1}
\newcommand{\dtwo}{\delta_2}
\newcommand{\dthree}{\delta_3}
\newcommand{\dfour}{\delta_4}

\newcommand{\dnot}{\delta_0}

\newcommand{\doned}{\delta_1(\delta)}
\newcommand{\dtwod}{\delta_2(\delta)}
\newcommand{\dthreed}{\delta_3(\delta)}
\newcommand{\dfourd}{\delta_4(\delta)}

\newcommand{\dnotd}{\delta_0(\delta)}

\newcommand{\fone}{f_1(\delta, \epsilon)}
\newcommand{\ftwo}{f_2(\delta, \epsilon)}

\newcommand{\tdftwo}{\td{f}_2(\delta, \epsilon)}
\newcommand{\gammad}{\gamma(\delta)}

\newcommand{\stoc}{\text{code with a stochastic encoder}}

\title{Arbitrarily Varying Remote Sources}
\author
{
  \IEEEauthorblockN{Amitalok J. Budkuley}
	\IEEEauthorblockA{CUHK\\
	amitalok@ie.cuhk.edu.hk
	}
	\and
	\IEEEauthorblockN{Bikash Kumar Dey}
	\IEEEauthorblockA{IIT Bombay\\
	bikash@ee.iitb.ac.in
	}
	\and
	\IEEEauthorblockN{Sidharth Jaggi}
	\IEEEauthorblockA{CUHK\\
	jaggi@ie.cuhk.edu.hk}
	\and
	\IEEEauthorblockN{Vinod M. Prabhakaran}
	\IEEEauthorblockA{TIFR Mumbai\\
	vinodmp@tifr.res.in}

}

\author{
  \IEEEauthorblockN{Amitalok J. Budkuley, Bikash Kumar Dey, Sidharth Jaggi and Vinod M. Prabhakaran}\\
  \IEEEauthorblockA{
    Emails: amitalok@ie.cuhk.edu.hk, bikash@ee.iitb.ac.in, jaggi@ie.cuhk.edu.hk, vinodmp@tifr.res.in }
}

\begin{document}
\maketitle 
\thispagestyle{empty}
\interdisplaylinepenalty=0
\begin{abstract}
We study a lossy source coding problem for an arbitrarily varying remote source (AVRS) which was proposed in a prior work. An AVRS transmits symbols, each generated in an independent and identically distributed manner,  which are sought to be estimated at the decoder. These symbols are \emph{remotely} generated, and the encoder and decoder observe noise corrupted versions received through a two-output noisy  channel. This channel is an arbitrarily varying channel controlled by a jamming adversary. We assume that the adversary knows the coding scheme as well as the source data non-causally, and hence, can employ malicious jamming strategies correlated to them. Our interest lies in studying the rate distortion function for codes with a stochastic encoder, i.e, when the encoder can \emph{privately} randomize while the decoder is \emph{deterministic}. We provide upper and lower bounds on this rate distortion function. %
\end{abstract}
%
%
\section{Introduction}\label{sec:introduction}
\begin{figure}[!ht]
  \begin{center}
    \includegraphics[trim=0cm 11cm 0cm 1cm, scale=0.34]{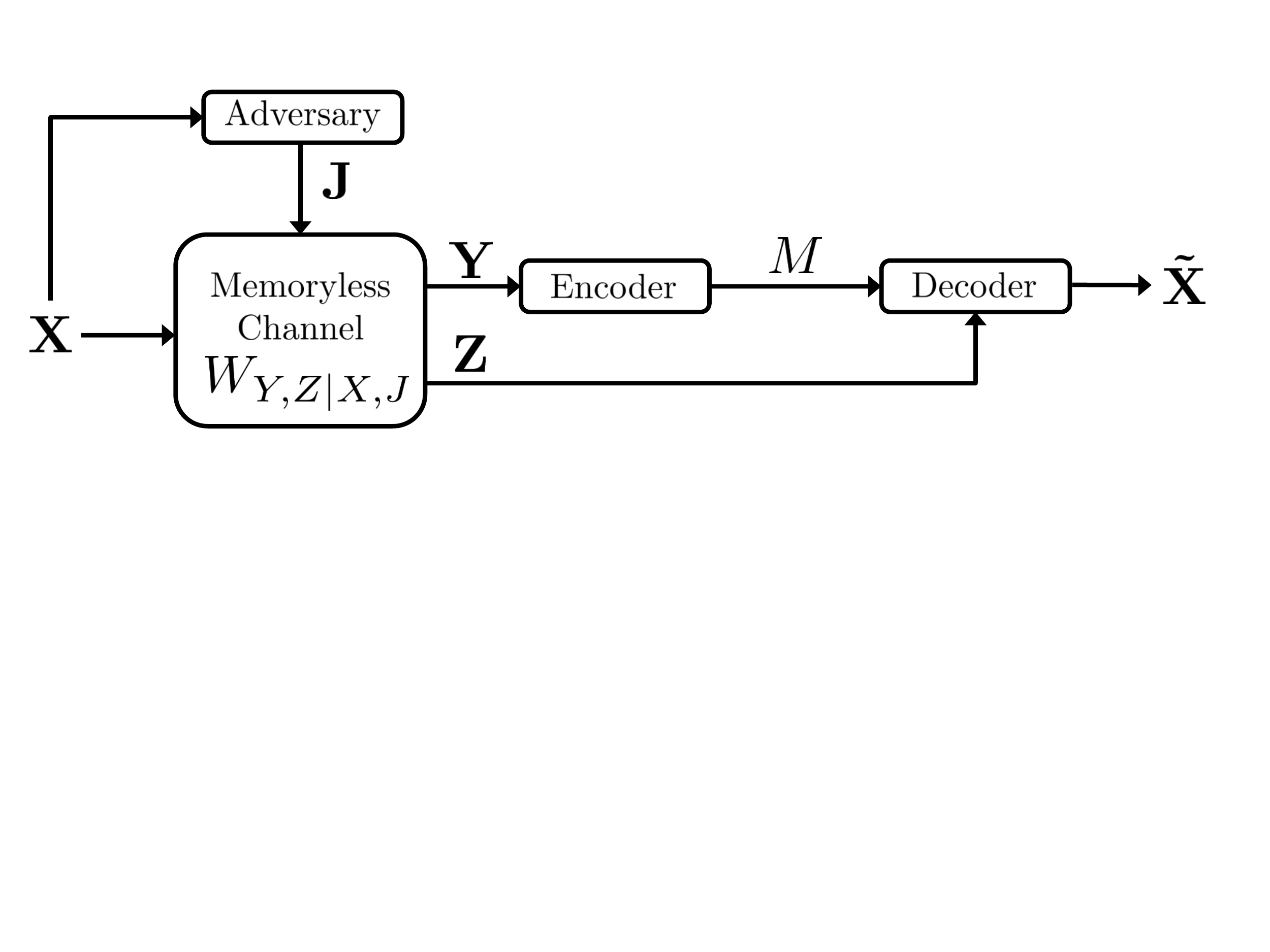}
    \caption{The arbitrarily varying remote source (AVRS) communication setup}
    \label{fig:main:setup:discrete}
  \end{center}
\end{figure}

The arbitrarily varying remote source (AVRS) model, depicted in Fig.~\ref{fig:main:setup:discrete}, was introduced in~\cite{bdp-isit2017}. 
Here a \emph{remote} source outputs a block of data $\bX$ comprising symbols generated in an  independent and identically distributed (i.i.d.) manner with a fixed distribution $P_X$. $\bX$ is observed over a noisy arbitrarily varying channel (AVC) $W_{Y,Z|X,J}$ partially controlled by an adversary's  jamming input $\bJ$. The two noisy versions $\bY$ and $\bZ$ of data $\bX$ are received as input at encoder and side information at decoder respectively. The encoder compresses $\bY$ into a message $M$ and transmits it losslessly to the decoder. The decoder then outputs an estimate $\btdX$ of $\bX$. The fidelity of the reconstruction is measured in terms of the average per letter distortion. The adversary knows the coding scheme and $\bX$, and hence, can leverage this information to design  pernicious jamming strategies. Our interest lies in the optimum rate of compression for a given a target distortion under any allowed jamming strategy. In this work, we study this problem when \emph{private} randomization at the encoder is allowed while the decoder is deterministic. 

Practical scenarios in some situations preclude the encoder's direct access to the exact source realization, unlike in standard lossy source coding~\cite{shannon-bstj1948}. For instance, a remote plant controller may be constrained to initiate control actions based upon a noisy view of plant variables. Dobrushin and Tsybakov~\cite{dobrushin-tsybakov-it1962} introduced the remote source coding problem, where the encoder observes the source through a fixed and known memoryless channel. 	The authors in~\cite{yamamoto-itoh-1980,draper-wornell-jsac2004} extended~\cite{dobrushin-tsybakov-it1962} to the scenario when the decoder too receives correlated side information, thereby generalizing the problem studied by Wyner and Ziv~\cite{wyner-ziv-it1976}. Unlike in these works where statistics are fixed, the AVRS models a robust scenario where an adversary can  induce arbitrary statistics   on the observations made at the encoder and decoder jointly. Furthermore, the AVRS model sits at the intersection of several other interesting lossy source coding problems. Apart from the aforementioned works, the AVRS model unifies  \emph{compound} and \emph{universal}  formulations of  several interesting problems (cf.~\cite{sakrison-inc1969,dembo-weissman-it2003,watanabe-etal-it2014}, and some of the references therein). In addition, some interesting adversarial source coding problems (cf.~\cite{berger-it1971,csiszar-korner-book2011,palaiyanur-it2011}) can also be modeled as special cases of the AVRS by making  appropriate assumptions on the AVC $W_{Y,Z|X,J}$ and the adversary.

It is well known that problems involving adversaries present challenges; results often crucially depend upon the nature of coding (possibility of randomization), adversary's  capabilities, error and/or distortion criteria etc. (see~\cite{lapidoth-narayan-it1998} for an excellent survey on problems involving AVCs). In our previous work~\cite{bdp-isit2017}, we studied an AVRS under \emph{randomized} coding where encoder-decoder share randomness unknown to the adversary. In that work, we gave upper and lower bounds on the randomized rate distortion function. Here we give results when randomization is restricted to the encoder only. Such codes, where the encoder can \emph{privately} randomize while decoding is \emph{deterministic}, are generally called \emph{codes with a stochastic encoder} (cf.~\cite{csiszar-korner-book2011}). Here is a summary of our  main contributions:
\begin{itemize}
\item We first extend our result in~\cite{bdp-isit2017} for the randomized rate distortion function  under the `average' (average over \emph{all} source sequences) distortion criteria, to a stricter `maximum' (maximum over \emph{all} typical source sequences) distortion criteria.
\item We use this `strengthening' of the result for randomized coding along with Ahlswede's elimination technique~\cite{ahlswede-1978} to extend our result for codes with a stochastic encoder, i.e., when private encoder-side randomization only is allowed.
\end{itemize}
Our proof of the upper bound under randomized coding is along  the lines of~\cite{bdp-isit2017}, which uses the \emph{refined} Markov lemma~\cite{bdp-it2017}, and involves careful modifications to the proof in~\cite{bdp-isit2017} necessitated by the stricter distortion criteria. To show our result for codes with a stochastic encoder, the aforementioned `strengthening' vis-\'a-vis the distortion metric is crucially used (see Remark~\ref{rem:elim}) in an intermediate de-randomization step, where we show the existence of a `good' randomized code with a polynomial-sized ensemble. 	

The rest of the paper is organized as follows. In
Section~\ref{sec:system:model}, we first introduce the notation and problem
setup. We state our main
results in Section~\ref{sec:main:results}. The proofs are presented in Sections~\ref{sec:proof} and~\ref{sec:proof:rd:obl:d}. We make concluding remarks in Section~\ref{sec:conclusion}.
\section{Notation and Problem Setup}\label{sec:system:model}
\subsection{Notation and Preliminaries}
Random variables are denoted by upper case letters (e.g. $X$), the values they
take by lower case letters (e.g. $x$) and their alphabets by calligraphic
letters (e.g. $\mathcal{X}$). 
We use boldface notation to denote random vectors (e.g. $\vec{X}$) and their values (e.g. $\vec{x}$). All vectors are of length $n$ (e.g. $\vec{X}=(X_1,X_2,\dots,X_n)$), where $n$ is the block length of operation. Also, we denote $\vec{X}^{i}=(X_1,X_2,\dots,X_i)$ and $\vec{x}^{i}=(x_1,x_2\dots,x_i)$ as well as  $\bX_{i}^{k}=(X_i,X_{i+1},\dots,X_k)$ and $\bx_{i}^{k}=(x_i,x_{i+1},\dots,x_k)$. We use the $l_{\infty}$ (denoted by $\|.\|_{\infty}$) norm for vectors.
Let $\cP(\mathcal{X})$ denote the set of all probability
distributions on a set $\mathcal{X}$. Similarly, let $\cP(\mathcal{X}|\mathcal{Y})$ be the set of all conditional distributions of a
random variable with alphabet $\mathcal{X}$ conditioned on another random
variable with alphabet $\mathcal{Y}$. For two random variables $X$ and $Y$,
we denote  the marginal distribution of $X$ obtained from the joint
distribution $P_{X,Y}$ by $[P_{X,Y}]_{X}$.  Distributions corresponding to
strategies adopted by the adversary are denoted by $Q$ instead of $P$ for
clarity. The set of all conditional distributions $\cP(\cJ|\cX)$ is
specifically denoted by $\sQ$. In cases where the subscripts are clear from the context, we sometimes
omit them to keep the notation simple. 
\removed{
Information quantities $H(X)$, $H(X,Y)$, $H(X|Y)$ and $I(X;Y)$ denote respectively the entropy of $X$, the joint entropy of $(X,Y)$, the conditional entropy of $X$ given $Y$ and the mutual information between $X$ and $Y$. 
}
Deterministic functions will be denoted in lowercase (e.g. $f$).  
We denote a
type of $X$ by $T_X$. Given sequences $\vec{x}$, $\vec{y}$, we
denote by $T_\vec{x}$ the type of $\vec{x}$, by $T_{\vec{x},\vec{y}}$ the joint
type of $(\vec{x},\vec{y})$ and by $T_{\vec{x}|\vec{y}}$ the conditional type
of $\vec{x}$ given $\vec{y}$. 
For $\epsilon\in(0,1)$, the set of $\epsilon$-typical set of $\vec{x}$ sequences for a distribution $P_X$ is
%
$\mathcal{T}^{n}_{\epsilon}(P_X)=\{\vec{x}:\|T_\vec{x}-P_X\|_{\infty}\leq \epsilon\}.$
%
In addition, for a joint distribution $P_{X,Y}$ and $\vec{x}\in\cX^n$, the conditionally typical set of $\vec{y}$ sequences, conditioned on $\vec{x}$, is defined as 
%
$\mathcal{T}^{n}_{\epsilon}(P_{X,Y}|\vec{x})=\{\vec{y}:\|T_{\vec{x},\vec{y}}-P_{X,Y}\|_{\infty}\leq \epsilon\}.$
%
\subsection{The Problem Setup}
Consider the communication setup depicted in Fig.~\ref{fig:main:setup:discrete}.
Let $\mathcal{X}$, $\mathcal{Y}$, $\mathcal{Z}$, $\mathcal{J}$ and
$\mathcal{\wtd{X}}$ denote finite sets. We consider a lossy source coding problem for an independent and identically distributed (i.i.d.) source with a distribution $P_X$ and alphabet $\cX$. We assume without loss of generality that $P_X(x)>0$, $\forall x\in\cX$. A length-$n$ block of data is sent over an arbitrarily varying channel (AVC). The AVC has two inputs, and two outputs. The two inputs comprise the data input $X\in\cX$ and adversary's jamming input $J\in\cJ$, while the two outputs are $Y\in\cY$ and $Z\in\cZ$.   We assume that the adversary knows $\bX$ non-causally and can randomize its jamming input $\bJ$. We denote its jamming strategy by $Q_{\bJ|\bX}$. The channel outputs $\bY$ and $\bZ$ are observed at the encoder and decoder respectively. The channel behaviour is given by the conditional distribution $W_{Y,Z|X,J}$. Thus, given inputs $\bx$ and $\bj$, channel outputs $\by$ and $\bz$ are observed with probability given by 
%
$\mathbb{P}\left( \vec{Y}=\vec{y},\vec{Z}=\vec{z}| \vec{X}=\vec{x},\bJ=\vec{j}\right)\hspace{-0.75mm}=\hspace{-0.75mm}\prod_{i=1}^n W_{Y,Z|X,J}(y_i,z_i|x_i,j_i).$
%
Upon observing $\bY$, the encoder compresses it to a message $M$ and send it losslessly to the decoder. Given $M$ and $\bZ$, the decoder then outputs an estimate $\btdX$ of the source data $\bX$. We assume that the adversary knows the coding scheme. The quality of the estimate is given in terms of the average per-letter distortion 
$d(\vec{X},\vec{\wtd{X}})=\frac{1}{n} \sum_{i=1}^n d(X_i,\wtd{X}_i),$
%
where $d:\mathcal{X}\times \mathcal{\wtd{X}}\rightarrow  \mathbb{R}^+$ denotes  a single-letter distortion measure with $d_{\max}=\max_{(x,\tilde{x})\in \mathcal{X}\times\mathcal{\wtd{X}}} d(x,\tilde{x})<\infty$.

An $(n,R)$ \emph{deterministic code} of block length $n$ and rate $R$ is a pair
$(\psi,\phi)$ of mappings, which consists of an encoder map $\psi:\cY^n\rightarrow
\{1,2,\dots,2^{nR}\},$ and a decoder map $\phi: \{1,2,\dots, 2^{nR} \}\times \cZ^n\rightarrow \ctdX^n.$ The encoder's output $M=\psi(\vec{Y})$ is received losslessly at the decoder. 
An $(n,R)$ \emph{randomized code} of block length $n$ and rate $R$ is a random variable which 
takes values in the set of all $(n,R)$ deterministic codes. 
We denote the encoder-decoder pair for this $(n,R)$
randomized code by $(\Psi,\Phi)$. This also forms the randomness $\Theta$ shared between the encoder-decoder, but unknown to the adversary. The message 
sent losslessly to the decoder is $M=\Psi(\vec{Y})$.
For this $(n,R)$ randomized code, the \ajblue{\emph{maximum expected distortion} or \emph{maximum distortion} $D^{(n)}$ is given by} 
\begin{equation}
D^{(n)}=\max_{\bx\in\cT^n_{\delta_0}(P_X)}\max_{Q_{\vec{J}|\vec{X}}} \mathbb{E}[d(\vec{x},\Phi(\Psi(\vec{Y}),\bZ))],\label{eq:D}
\end{equation}
where the expectation is over the channel and the adversary's jamming action and the shared randomness. 
Note that the first maximization above is over source sequences $\bx$ which are $\delta_0$-typical according to distribution $P_X$. Here $\delta_0=\delta_0(n)$ is a fixed sequence which depends on $n$. Our distortion criterion here differs from the \emph{usual} average distortion criterion (cf.~\cite{elgamal-kim}) 
\begin{IEEEeqnarray*}{rCl}
D^{(n)}=\max_{Q_{\vec{J}|\vec{X}}} \mathbb{E}[d(\vec{X},\Phi(\Psi(\vec{Y}),\bZ))],\yesnumber\label{eq:D:avg}
\end{IEEEeqnarray*}
where distortion is   averaged over all $\bx$ sequences under the i.i.d. distribution $P_X$.  Our earlier work in~\cite{bdp-isit2017} considered this criterion. Note that the criterion in~\eqref{eq:D}, which has also been considered in other works (cf.~\cite{csiszar-korner-book2011}), is \emph{stronger} than the one in~\eqref{eq:D:avg}, and proves crucial in our proof of Theorem~\ref{thm:main:result:s} (see Remark~\ref{rem:elim}). In a manner similar to~\cite{csiszar-korner-book2011}, we assume here that $\delta_0(n)\rightarrow 0$ and $\sqrt{n}\delta_0(n)\rightarrow \infty$ as $n\rightarrow \infty$. To keep the notation simple, we henceforth suppress the dependence of $\delta_0$ on $n$.  

An $(n,R)$ \emph{code with a stochastic encoder} of block length $n$ and rate $R$ is a pair
$(\Psi,\phi)$ of mappings consisting of a \emph{randomized} encoding map $\Psi:\cY^n\rightarrow
\cP(\{1,2,\dots,2^{nR}\}),$ and a \emph{deterministic} decoding map $\phi: \{1,2,\dots, 2^{nR} \}\times\cZ^n\rightarrow \ctdX^n.$ The encoder's output message $M=\Psi(\vec{Y})$ is received error-free at the decoder. 
For this $(n,R)$ \stoc, the \emph{maximum distortion} $D^{(n)}$ is given by 
\begin{equation}\label{eq:D:stoc}
D^{(n)}=\max_{\bx\in\cT^n_{\delta_0}(P_X)}\max_{Q_{\vec{J}|\vec{X}}} \mathbb{E}[d(\vec{x},\phi(\Psi(\vec{Y}),\bZ))],
\end{equation}
where the expectation is over the channel, the adversary's jamming action and the encoding. 

Given a target distortion $D$, a rate  $R$ is \emph{achievable} under randomized coding if for any $\epsilon>0$ there exists an $n_0(\epsilon)$ such that for every $n\geq n_0(\epsilon)$ there exists an $(n,R)$ randomized code with the resulting \ajblue{maximum distortion} $D^{(n)}\leq D+\epsilon$. We define the adversarial rate distortion function under randomized coding $R_r(D)$ as the infimum of all achievable rates. These definitions for achievable rate and rate distortion function can be analogously stated for codes with  a stochastic encoder. We denote by $R_s(D)$ the rate distortion function for codes with a stochastic encoder. Our aim in this work is to obtain upper and lower bounds on the adversarial rate distortion functions under randomized coding as well as  for codes with a stochastic encoder under the distortion criteria stated in~\eqref{eq:D} and~\eqref{eq:D:stoc} respectively.
\section{The Main Result}\label{sec:main:results}
%
%
Recall that $\sQ= \cP(\cJ|\cX)$  denotes the set of all conditional
distributions of $J$ given $X$.
For any distribution $Q_{J|X}\in\sQ$,
consider the single-letter joint distribution
$P_XQ_{J|X}W_{Y,Z|X,J}$. 
Let $D_0 :=  \min_{P_{\tilde{X}|Y,Z}}\max_{Q_{J|X}\in \sQ} \mathbb{E}[d(X,\tilde{X})]$ and $D_1 :=  \min_{P_{\tilde{X}|Z}}\max_{Q_{J|X}\in \sQ} \mathbb{E}[d(X,\tilde{X})]$.
\removed{
\begin{align}
D_0 &=  \min_{P_{\tilde{X}|Y,Z}(\cdot,\cdot)} \ \max_{Q_{J|X}\in \sQ} \mathbb{E}[d(X,\tilde{X}(Y,Z))]\label{eq:d0}\\
\text{and}\qquad D_1 &=  \min_{P_{\tilde{X}|Z}}\max_{Q_{J|X}\in \sQ} \mathbb{E}[d(X,\tilde{X}(Z))].\label{eq:d1}
\end{align}
}
Here $D_0$ is the minimax \ajblue{expected distortion}
when $\bX$ is jointly estimated from $\bY$ and $\bZ$, while $D_1$ is the minimax \ajblue{expected distortion when $\bX$} is estimated from $\bZ$ only.
Consider an auxiliary random variable $U$ with a 
finite alphabet $\mathcal{U}$ distributed according to $P_{U|Y}$,
such that $(X,J,Z) \leftrightarrow Y \leftrightarrow  U$ forms a Markov chain.
The joint distribution of $(X,J,Y,Z,U)$ is then given by
$P_XQ_{J|X}W_{Y,Z|X,J}P_{U|Y}$. 
Let us now define the following.
\begin{align}\label{eq:r:d:u}
R_U^*(D):=
\begin{cases}
 \displaystyle\min_{ P_{U|Y},\ \zeta(\cdot,\cdot)    } \ \max_{Q_{J|X}\in \sQ  }I(U;Y|Z),\hspace{-2mm}& \mbox{if } D\in[D_0,D_1]\\
\hspace{5mm}0, &\mbox{if } D> D_1.
\end{cases}
\end{align}
The minimization above is over $P_{U|Y}\in\mathcal{P}(\mathcal{U}|\mathcal{Y})$ and 
$\zeta:\mathcal{U}\times\mathcal{Z}\rightarrow \wtd{\mathcal{X}}$ such that
 $\mathbb{E}[d(X,\tilde{X})]\leq D,~\forall Q_{J|X}\in \sQ$. Note that the cardinality $|\cU|$ of $\cU$ can be restricted to $|\cU|\leq |\ctdX|^{|\cZ|}$, which is the number of possible functions from $\cZ$ to $\ctdX$.
\begin{align}\label{eq:r:d:l}
R_L^*(D):=
\begin{cases}
 \displaystyle\max_{Q_{J|X}\in \sQ  } \ \min_{  P_{U|Y},\ \zeta(\cdot,\cdot)   } I(U;Y|Z), \hspace{-2mm}&\mbox{if } D\in[D_0,D_1] \\
\hspace{5mm}0, &  \mbox{if } D> D_1,
\end{cases}
\end{align}
where the minimization is over $P_{U|Y}\in\mathcal{P}(\mathcal{U}|\mathcal{Y})$
and $\zeta:\mathcal{U}\times\mathcal{Z}\rightarrow \wtd{\mathcal{X}}$   such that $\mathbb{E}[d(X,\tilde{X})]\leq D$ for the specified $Q_{J|X}$. In a manner similar to~\cite{wyner-ziv-it1976}, we can restrict the cardinality of $U$ to $|\cU|\leq |\cY|+1$.
Next, we state our results.
%
\begin{theorem}\label{thm:main:result:dmc}
The adversarial rate distortion function  $R_r(D)$ under randomized coding is such that
\begin{equation*}
R_L^*(D)\leq R_r(D)\leq  R_U^*(D).\label{eq:R:D:r}
\end{equation*}
\end{theorem}
The proof uses the approach in~\cite{bdp-isit2017}; see Section~\ref{sec:proof} for details. 
We now state our main result which shows that the adversarial rate distortion function under codes with a stochastic encoder equals that under randomized codes.\\
%
\begin{theorem}\label{thm:main:result:s}
The adversarial rate distortion function under codes with a stochastic encoder is equal to that under randomized coding, i.e., 
\begin{equation*}
R_s(D)=R_r(D),
\end{equation*}
and hence,  %
\begin{equation*}
R_L^*(D)\leq R_s(D)\leq  R_U^*(D).\label{eq:R:D:s}
\end{equation*}
\end{theorem}
The proof is given in~Section~\ref{sec:proof:rd:obl:d}. 
\begin{remark}
The results in Theorem~\ref{thm:main:result:s} continue to hold  for codes with a stochastic encoder under the `usual' average distortion criterion. Note that for this criterion, we replace the maximum (over all $\delta_0$-typical $\bx$ sequences) in~\eqref{eq:D:stoc}  by an average (over all $\bx$ sequences). The converse directly follows from that in Theorem~\ref{thm:main:result:s}. The assertion now follows by noting that  our achievability proof under the `stronger' criterion in~\eqref{eq:D:stoc} guarantees achievability under the `average' distortion criterion.  
\end{remark}
\section{Proof of Theorem~\ref{thm:main:result:dmc}}\label{sec:proof}
\subsection{Achievability} \label{sec:achieve} We use the approach in~\cite{bdp-isit2017,bdp-arxiv2017}. 
Note that while the analysis proceeds along similar lines, there are important differences. These result from the `stricter' requirement of guaranteeing that $\mathbb{E}[d(\bx,\btdX)]\leq D+\epsilon$, $\forall \bx\in\cT^n_{\delta_0}(P_X)$, unlike in~\cite{bdp-arxiv2017}, where we required $\mathbb{E}[d(\bX,\btdX)]\leq D+\epsilon$. In particular, we establish more general versions of several claims in~\cite{bdp-arxiv2017} by careful modifications (see, for instance, Claim~\ref{claim:tdx} and other associated claims where, unlike in~\cite{bdp-arxiv2017}, there is no averaging over $P_X$ and the vector $\bx\in\cT^n_{\delta_0}(P_X)$ is now fixed) necessitated by the aforementioned requirement. 

We first present a brief outline of the coding scheme and the analysis. The detailed proof can be found in Appendix~\ref{app:ach:proof}.
%
%
First, note that for $D>D_1$, we can directly estimate $\bX$ from the side information $\bZ$ using a possibly randomized estimator $P_{\tilde{X}|Z}$. Hence, we have $R(D)=0$ for $D>D_1$. Let us fix $D_1\geq D \geq D_0$. 
We now fix  $P_{U|Y}$ and 
\ajblue{$\zeta(u,z)$ given} by~\eqref{eq:r:d:u}, and prove the achievability of the rate
\begin{align*}
\ajblue{R^{(P_{U|Y},\zeta)} } := \max_{Q_{J|X}\in \sQ}(I(U;Y)-I(U;Z)).
\end{align*}
Following the analysis in~\cite{bdp-isit2017}, we can express $R^{(P_{U|Y},\zeta)}$ as\footnote{Here we indicate $I(U;Y)$ as a function
of only $P_Y$ as in our proof of achievability we have fixed $P_{U|Y}$. For the same reason,
we indicate $I(U;Z)$ only as a function of $Q_{J|X}$, as the other distributions 
$P_X, P_{U|Y}$, and  $W_{Y,Z|X,J}$ are fixed in our discussion.\\
By the notation $P_Y' \stackrel{f(\epsilon)}{\approx} P_Y$, we mean  that 
$||P_Y' - P_Y||_\infty \leq f(\epsilon)$ for an appropriate $f(\epsilon)>0$, where $f(\epsilon)\rightarrow 0$ as $\epsilon\rightarrow 0$. See~\cite{bdp-isit2017} for details. 
}
\begin{align}
R^{(P_{U|Y},\zeta)} \geq \max_{P'_Y \in \cP(\cY)}\hspace{-0.5mm} \left[I_{P'_Y}(U;Y)- \hspace{-1.5mm}\min_{\begin{array}{c}Q_{J|X}\in \sQ \\ P_Y \stackrel{f(\epsilon)}{\approx} P'_Y\end{array}}\hspace{-1.5mm} I_{Q_{J|X}}(U;Z)\right] - \frac{\epsilon}{4}. \label{eq:rds}
\end{align}
Let us now define for every type $T_Y\in\cP(\cY)$, 
\begin{IEEEeqnarray}{rCl}
R_U (T_Y) & :=& I_{T_Y}(U;Y)+ \frac{\epsilon}{4} \label{eq:ru}\\
\tilde{R}(T_Y)& :=& \min_{\begin{array}{c} Q_{J|X}\in \sQ\\ P_Y \stackrel{f(\epsilon)}{\approx} T_Y \end{array}} \hspace{-1.5mm}I_{Q_{J|X}}(U;Z) - \frac{\epsilon}{4}.
\label{eq:rtilde}
\end{IEEEeqnarray}%
\emph{Code Construction:}
\begin{itemize}
\item The random code generation is as follows. Here we assume
that the entire ensemble of all possible codes is shared between the encoder and decoder, and they jointly select, at random, a code from the ensemble using the shared randomness $\Theta$. Note that this process is equivalent to generating the code randomly and then sharing it with the encoder-decoder.
\item Fix type $T_Y\in\cP(\cY)$. Generate a codebook $\cC(T_Y)$ with $2^{nR_U (T_Y)}$ vectors i.i.d. $\sim P_U$, where $P_U := \left[T_YP_{U|Y}\right]_U$. $\cC(T_Y)$ is randomly partitioned into $2^{n(R_U (T_Y)-\tilde{R}(T_Y))}$ bins. Do this for every $T_Y\in\cP(\cY)$. 
\item We share between the encoder and decoder the entire list of binned codebooks for all $T_Y\in\cP(\cY)$. 
\end{itemize}
\emph{Encoder operations:}
\begin{itemize}
\item The encoder observes $\by$ and computes its type $T_{\by}$.
It finds if there exists at least one codeword in $\cC(T_{\by})$ which is jointly typical with $\by$ with respect to (w.r.t.) the distribution $T_{\by}P_{U|Y}$.
If there exists at least one possible codeword, it selects one from amongst them at random and sends its bin index in $\cC(T_{\by})$ along with $T_{\by}$ to the decoder.
\item Since the number of types are polynomial in $n$, for large enough $n$, the rate required to convey $T_{\by}$ is at most
$\epsilon/4$. Hence, the rate of the entire message is bounded by 
\begin{align*}
R 
& \leq \max_{T_Y} (R_U (T_Y)-\tilde{R}(T_Y)) +\frac{\epsilon}{4}\\
& \leq  R^{(P_{U|Y},\zeta)}+\epsilon.   \hspace{10mm} \text{(using \eqref{eq:rds},~\eqref{eq:ru} and~\eqref{eq:rtilde})}
\end{align*}
\end{itemize}
\emph{Decoder operations:}
\begin{itemize}
\item The decoder observes side information $\bZ=\bz$, and receives $T_{\by}$ and the bin index. It first identifies the following set $\sQ^{(n)}(T_\by)$ of valid  conditional types corresponding to block length $n$ 
\begin{align*}
\sQ^{(n)}(T_\by) & := \{T_{J|X}\in \sQ : [P_X T_{J|X}W_{Y,Z|X,J}]_Y \hspace{0mm} \stackrel{f(\epsilon)}{\approx} T_\by\}.
\end{align*}
Here every $T_{J|X}\in \sQ^{(n)}(T_\by)$ induces a $Y$-marginal distribution  `close' to $T_{\by}$.
\item 
Next, the decoder checks within the bin if there is a codeword $\bu\in\cC(T_{\by})$ such that $(\bu,\bz)$ is jointly typical w.r.t. the distribution 
$\left[P_XT_{J|X}W_{Y,Z|X,J}P_{U|Y}\right]_{U,Z}$ for some $T_{J|X} \in \sQ^{(n)}(T_{\by})$. 
It chooses that  codeword if unique, otherwise it chooses an arbitrary codeword $\bu$ from the bin. It then outputs $\td{\bx}$, where  $\td{x}_i=\zeta(u_i,z_i)$, $i=1,2,\dots,n$, using the chosen codeword $\bu$ and $\bz$.
\end{itemize}
\ajblue{ \emph{Maximum distortion analysis:}}
\begin{itemize}
\item We fix a source sequence $\bx\in\cT^n_{\delta0}(P_X)$, and show that under any feasible jamming strategy  $Q_{\bJ|\bX=\bx}$, the resulting distortion is at most $D+\epsilon$. 
As this distortion does not depend on the chosen sequence $\bx\in\cT^n_{\delta_0}(P_X)$, it follows that the overall distortion $D^{(n)}\leq D+\epsilon$, where $D^{(n)}$ is given by~\eqref{eq:D}. The detailed proof follows the approach in~\cite{bdp-isit2017} and uses the refined Markov lemma~\cite[Lemma~8]{bdp-isit2017}. Refer Appendix~\ref{app:ach:proof} for details.
\end{itemize}

\subsection{The proof of the lower bound} Our converse follows from the converse in~\cite{bdp-isit2017}. There it is shown that any achievable rate is lower bounded by
the maximin expression in~\eqref{eq:r:d:l}, though under the usual distortion criterion given in~\eqref{eq:D}. However, the distortion criterion in~\eqref{eq:D:avg} is a `stronger' criterion as a rate $R$ which is not achievable under~\eqref{eq:D:avg} will also not be achievable under~\eqref{eq:D}.  
\section{Proof of Theorem~\ref{thm:main:result:s}}\label{sec:proof:rd:obl:d}
Clearly, $R_s(D)\geq R_r(D)$. 
We now give the proof of achievability to show that $R_s(D)= R_r(D)$; the bounds on $R_s(D)$ then directly follow from those given for $R_r(D)$ in Theorem~\ref{thm:main:result:dmc}. This proof uses the approach in~\cite{ahlswede-1978} and  has two parts, viz., part (a) and part (b). In part (a), we show that given any randomized code of rate $R\geq R_s(D)$,  there exists a randomized code  with the same rate $R$ but with an ensemble size $n^2$. In part $(b)$, we construct a code with a stochastic encoder with rate arbitrarily close to $R$, thereby completing the proof.\\
\noindent {\it Part (a)}: 
%
Let $\epsilon>0$. Consider any randomized code, say $\cC=(\Psi,\Phi)$, of rate $R\geq R_r(D)$ for which the resulting \ajblue{maximum distortion} is $D^{(n)}\leq D+\epsilon$. 
Consider $K$ independent repetitions of a random experiment of deterministic codebook selection from the randomized code $\cC$. We denote the $K$ outcomes, i.e., deterministic codes, by $C_i:=(\psi_i,\phi_i)$, $i=1,2,\dots, K$, where $(\psi_i,\phi_i)$ denote the encoder-decoder  pair for code $C_i$. Given any $\bx\in\cT^n_{\delta_0}(P_X)$ and jamming input $\bj\in\cJ^n$, let the corresponding distortion under code $C_i$ be $D^{(n)}(\bx, \bj, C_i):= \mathbb{E}[d(\bx,\btdX)]$, where the expectation is over the channel $W_{Y,Z|X,J}$. Note that $\forall \bx\in\cT^n_{\delta_0}(P_X)$ and $\forall \bj\in\cJ^n$, we have
\begin{IEEEeqnarray*}{rCl}
\bbE_{\cC}[D^{(n)}(\bx, \bj, C_i)]\leq D+\epsilon. \nonumber\label{eq:avg:tdD}
\end{IEEEeqnarray*}
%
We now state the following useful result. 
\begin{lemma}[{~\cite[pg.~16]{ahlswede-1980}}]\label{lem:bernstein}
Let $Z_i$, $i=1,2,\dots,N$, be a sequence of discrete independent random variables that take values in $[-b,b]$, \ajblue{where $b\in(0,\infty)$.} Then, for any $\mu>0$, there exists $0<\alpha\leq \min{\{1,\frac{b}{2}e^{-2b}\}}$ such that
\begin{IEEEeqnarray*}{rCl}
\mathbb{P}\left(\frac{1}{N}\sum_{i=1}^N (Z_i -\bbE[Z_i])\geq \mu  \right)\leq e^{-\left( \alpha \mu+\alpha^2b^2\right)N}.
\end{IEEEeqnarray*}
\end{lemma}
Observe that $D^{(n)}(\bx,\bj,C_i)$, $i=1,2,\dots, K$ are i.i.d.. Further, $\forall i$, $D^{(n)}(\bx,\bj,C_i)\in [0, d_{\max}]$, where $d_{\max}<\infty$. Hence, $D^{(n)}(\bx,\bj,C_i)$, $\forall i$  are bounded. Let us define 
\begin{IEEEeqnarray}{rCl}
\bbD^{(n)}(\bx,\bj,C_i):=D^{(n)}(\bx,\bj,C_i)-\bbE[D^{(n)}(\bx,\bj,C_i)].\label{eq:new:D}
\end{IEEEeqnarray} 
%
Now for any $\mu>0$, there exists $\alpha\in(0,\min{\{1,\frac{b}{2}e^{-2b}\}}]$, \ajblue{where $b:=d_{\max}\in(0,\infty)$, such that}
\begin{IEEEeqnarray*}{rCl}
\bbP_{\cC}&\left(\frac{1}{K} \sum_{i=1}^K  \bbD^{(n)}(\bx,\bj,C_i)\geq \mu  \right)\stackrel{}{\leq}  e^{-\left( \alpha \mu+\alpha^2b^2\right)K}, \nonumber\label{eq:P1}
\end{IEEEeqnarray*}
where we have used~\eqref{eq:new:D} and Lemma~\ref{lem:bernstein}.
Now from the union bound, we get
\begin{IEEEeqnarray*}{rCl}
&\mathbb{P}_{\cC}&\left(\exists \bx\in\cT^n_{\delta_0}(P_X), \bj\in\cJ^n: \frac{1}{K} \sum_{i=1}^K \bbD^{(n)}(\bx,\bj,C_i)\geq \mu \right)\\
&\stackrel{}{\leq} & |\cX|^n |\cJ|^n e^{-\left( \alpha \mu+\alpha^2b^2\right)K}\\
&=&  e^{-(K\left( \alpha \mu+\alpha^2b^2\right)-n\log(|\cJ||\cX|))},\yesnumber\label{eq:union}
\end{IEEEeqnarray*}
which is vanishing as $n\rightarrow \infty$ when $K=n^2.$
Thus, for  $\epsilon>0$, there exists a randomized code with ensemble size $K=n^2$ such that  for $n$ sufficiently large, the corresponding expected distortion $D^{(n)}(\bx,\bj)\leq D+\epsilon$, $\forall\bx\in\cT^n_{\delta_0}(P_X)$ and $\bj\in\cJ^n$. As $\epsilon>0$ is arbitrary, this completes the proof of this part.
\begin{remark}\label{rem:elim}
The choice of the distortion criteria in~\eqref{eq:D} is crucial for the application of the elimination technique~\cite{ahlswede-1978} in our proof above. If the `usual' distortion criterion, viz., the  average (averaged over source) distortion criterion~\eqref{eq:D:avg}, had been chosen, then it would have  necessitated taking a union bound in~\eqref{eq:union} over all functions $f:\cX^n\rightarrow \cJ^n$,  instead of pairs of sequences $(\bx,\bj)$. However, the number of such functions $f:\cX^n\rightarrow \cJ^n$ grows as $|\cJ^n|^{|\cX^n|}$, i.e., doubly-exponentially  in $n$. Thus, guaranteeing a vanishing probability of error in~\eqref{eq:union} under a polynomial-sized code collection would not have been possible using this approach. 
\end{remark}
\noindent{\it Part (b):} 
For this part, let us first denote the  randomized code of ensemble size $n^2$ in part $(a)$ by $(\Psi,\Phi):=\{\psi_k,\phi_k\}$, where $k=1,2,\dots,n^2$.  Recall that the rate of this code $(\Psi,\Phi)$ is $R\geq R_r(D)$, and given any $\epsilon>0$ and $n$ large enough,  its 
corresponding \ajblue{maximum distortion} $D^{(n)}\leq D+\epsilon$. 
For our code with a stochastic encoder, denoted by $(\td{\Psi},\td{\phi})$, the encoder first chooses, uniformly at random and \emph{privately}, a deterministic code, say $(\psi_I,\phi_I)$ from the $n^2$-sized ensemble of code $(\Psi,\Phi)$. The encoder then sends this index $I\in[1,2,\dots, n^2]$  to the decoder using $\log(n^2)$ bits, followed by the corresponding codeword $\psi_I(\by)$. The \emph{informed} decoder then outputs the estimate $\phi_I(\psi_I(\by))$. Note that the overall distortion of this code $(\td{\Psi},\td{\phi})$ constructed from $(\Psi,\Phi)$ is $D+\epsilon$. The rate $\td{R}$ of this code $(\td{\Psi},\td{\phi})$ is $\td{R}=R+\frac{\log(n^2)}{n}$. However, as $\frac{2\log(n)}{n}\rightarrow 0$ as $n\rightarrow \infty$, we have $\td{R}\rightarrow R$ as $n\rightarrow \infty$. Thus, by choosing a sufficiently large  $n$, we get a code with a stochastic encoder with rate arbitrarily close to $R$ and \emph{maximum  distortion} $D^{(n)}\leq D+\epsilon$. This completes the proof of part $(b)$, and thus, the proof of the theorem.
\section{Conclusion}\label{sec:conclusion}We studied lossy source coding for an arbitrarily varying remote source. Here statistics of the noisy source observations at the encoder and decoder are controlled by an adversary and vary arbitrarily across time. The adversary knows the coding scheme and the source data non-causally, and hence, can employ malicious strategies. We studied the rate distortion function when  the encoder can privately randomize, i.e., for codes with a stochastic encoder. Toward this, we first extended an earlier result for randomized coding under the \emph{usual} average (averaged over all sequences) distortion criterion to a `stronger' maximum (over all typical source sequences) distortion criterion. Using this `strengthening' of the result under randomized coding, we then showed that the result remains unchanged if randomization is restricted to the encoder only. 
\section*{Acknowledgments}
	A.~J.~Budkuley acknowledges several helpful discussions with S.Vatedka (CUHK) and M. Bakshi (CUHK). A.~J.~Budkuley and S.~Jaggi were supported in part by a grant from the University Grants  Committee of the Hong Kong Special Administrative Region (Project No.\  AoE/E-02/08) and RGC GRF grants 14208315 and 14313116. B.~K.~Dey was supported in part by Bharti Center for Communication, IIT Bombay.

\appendices
\section{Proof of Achievability}\label{app:ach:proof}
In this detailed proof of achievability, we begin with the description of our randomized coding scheme.\\
\noindent \emph{Code Construction:}
\begin{itemize}
\item As discussed in the outline, the random code $\cC$ is a list of
individual codes $\cC(T_Y)$ for every type $T_{Y}\in\sT^{\nb}(\cY)$. This
list of codes is
shared as the common randomness $\Theta$ between the encoder-decoder.  
\item
For a fixed type $T_Y\in\sT^{\nb}(\cY)$, our code $\cC(T_Y)$ is a binned codebook
comprising $2^{n R_U(T_Y)}=2^{n(R(T_Y)+\tilde{R}(T_Y))}$ vectors
$\vec{U}_{j,k}$, where $j=1,2,\dots,2^{nR(T_Y)}$ and
$k=1,2,\dots,2^{n\tilde{R}(T_Y)}$. Here $R_U(T_Y)$ and $\tilde{R}(T_Y)$
are as given in \eqref{eq:ru} and \eqref{eq:rtilde} respectively, and
$R(T_Y) = R_U(T_Y) - \tilde{R}(T_Y)$. Every
codeword $\vec{U}_{j,k}$ is chosen i.i.d. $\sim P_U$, where $P_U:=[P_{U|Y}
T_{Y}]_{U}$. There are $2^{nR(T_Y)}$ bins indexed
by  $j$, with each bin containing $2^{n\tilde{R}(T_Y)}$ codewords indexed by
$k$. Let $\cB^{(T_Y)}_m$ denote the bin with index $m$.
Thus, our
code $\cC$ is the list containing $\cC(T_Y); T_Y\in\sT^{\nb}(\cY)$.
\end{itemize}
\emph{Encoding:}
\begin{itemize}
\item Given input $\bY$, the encoder determines its type $T_{\bY}$ to identify
$\cC(T_{\bY})$. In $\cC(T_{\bY})$, it finds a codeword  $\vec{U}_{m,l}$, where $m\in
\{1,2,\dots,2^{nR^{(T_{\bY})}}\}$ and $l\in
\{1,2,\dots,2^{n\tilde{R^{(T_{\bY})}}}\}$, such that 
\begin{equation}\label{eq:encoder:condition:dmc}
\|T_{\vec{U}_{m,l},\vec{Y}}-P_{U|Y}T_{\bY}\|_{\infty}\leq \delta_2(\delta).
\end{equation}
Here $\delta_2(\delta)>0$ is a fixed constant (the choice of 
$\delta_2(\delta)$ is indicated in Lemma~\ref{lem:covering})\footnote{Here $\delta>0$ is a function of $\epsilon$, such that
$\delta\rightarrow 0$ as $\epsilon \rightarrow 0$ and it is such that
\eqref{eq:D:eps} holds.}. This implies
that $\vec{U}_{m,l}$ and $\vec{Y}$ are jointly
typical according to the distribution $P_{U|Y} T_{\bY}$. If no such
$\vec{U}_{m,l}$ is found, then the encoder chooses $\bU_{1,1}$. If more than
one $\vec{U}_{m,l}$ satisfying~\eqref{eq:encoder:condition:dmc} exist, then the
encoder chooses one uniformly at random from amongst them. Let
$\vec{U}=\vec{U}_{M,L}$ denote the chosen codeword. 
\item The encoder transmits $T_{\bY}$ and the bin index $M$ losslessly to the decoder. 
\end{itemize}
\emph{Decoding:}
\begin{itemize}
\item Let the bin index $m$ and side information $\vec{z}$ be received at the
decoder. In addition, the decoder knows the type $T_{\by}$ of
the encoder's input $\by$, and so the
code $\cC(T_{\by})$ used by the encoder. 
\item For some fixed parameter $\gamma(\delta)>0$
(the choice of $\gamma(\delta)$ is indicated in Lemma~\ref{lem:u:in:L}), the decoder determines the set of
codewords 
\begin{IEEEeqnarray}{rCl}\label{eq:dec:cond}
\mathcal{L}_{\gamma(\delta)}(m,\vec{z})=\Big\{ \vec{u}\in \mathcal{B}^{(T_{\by})}_m&: \|T_{\vec{u},\vec{z}}-[P_X T_{J|X} W_{Y,Z|X,J} P_{U|Y}]_{U,Z}\|_{\infty}\leq\gamma(\delta), \text{ for some } T_{J|X}\in \sQ(T_{\by}) \Big\}, 
\end{IEEEeqnarray}
%
Here $\sQ(T_{\by}):=\{T_{J|X}\in\sT^n(\cJ|\cX): [P_X T_{J|X} W_{Y,Z|X,J}]_Y\stackrel{f(\epsilon)}{\approx} T_{\by} \}$, where recall that by the notation $P_Y' \stackrel{f(\epsilon)}{\approx} P_Y$, we mean  that 
$||P_Y' - P_Y||_\infty \leq f(\epsilon)$ for an appropriate $f(\epsilon)>0$, where $f(\epsilon)\rightarrow 0$ as $\epsilon\rightarrow 0$.  
%
%
%
\item If $\mathcal{L}_{\gamma(\delta)}(m,\vec{z})$ contains exactly one
codeword, then the decoder chooses it. Otherwise it chooses $\bu_{m,1}$. Let the
chosen codeword be $\vec{u}_{m,\tilde{l}}$. 
\item The decoder outputs $\td{\bx}$, where $\td{x}_i=\eta(u_i(m,\tilde{l}),z_i)$. 
\end{itemize}
\emph{Maximum distortion analysis:}
%
To begin, let us fix $\bx\in\cT^n_{\delta_0}(P_X)$. We first analyse the error in decoding the codeword $\bU=\bU_{M,L}$ chosen by the encoder. 
The decoder makes an error if one or more of the following events occur.
\begin{IEEEeqnarray*}{rCl}\label{eq:err:events}
E_{\text{enc}}&=&\{(\bU_{j,k},\bY)\not \in \cT^n_{\dtwo} (P_{U|Y} T_{\bY}), \forall j,k\}	\\
E_{\text{dec},1}&=&\{(\bU,\bZ)\not \in \cL_{\gamma(\delta)}(M,\bZ) \}	\\
E_{\text{dec},2}&=&\{(\bU_{M,l'},\bZ) \in \cL_{\gamma(\delta)}(M,\bZ) \text{ for some } l'\neq L\}	,
\end{IEEEeqnarray*}
Then, using the union bound we can express the probability of decoding error by
\begin{IEEEeqnarray}{rCl}\label{eq:P:E}
\bbP(E)\leq \bbP(E_{\text{enc}})+\bbP(E_{\text{dec},1}|E^c_{\text{enc}})+\bbP(E_{\text{dec},2}|E^c_{\text{enc}}).
\end{IEEEeqnarray}
We will show that for every $\epsilon>0$ there exists small enough $\delta>0$ 
such that $\bbP(E)\rightarrow 0$ as $n\rightarrow \infty$.
We first make the following obvious claim. 
\begin{claim}
Let $\bU$ be generated i.i.d. $\sim P_U$. Then, with probability at least $(1 - |\cU|e^{-2n\delta^2})$, $\bU \in 
\cT^n_{\delta}(P_U)$.
\end{claim}
Let us define this ``good'' event as $A_{U}:=\{\bU \in \cT^n_{\delta}(P_U)\}$.
We now state the following lemma which guarantees that the first term in~\eqref{eq:P:E} is vanishingly small.
\begin{lemma}\label{lem:covering}
Under the event $A_U$, there exist $\dtwod,\fone>0$, where $\dtwod,\fone\rightarrow 0$
as $\delta,\epsilon\rightarrow 0$,
such that the encoder
finds a codeword $\bU$ with probability at least $1-2^{-2^{n\fone}}$ such that
$(\bY,\bU)\in\cT^n_{\dtwo}(P_{U|Y}T_{\bY})$. 
\end{lemma}
The proof of this lemma follows from the covering lemma~\cite[Lemma~3.3]{elgamal-kim}. 
Note that this lemma specifies the $\dtwod$ parameter which appears in the definition of the encoder in~\eqref{eq:encoder:condition:dmc}.
This lemma implies $\bbP(E_{\text{enc}}) \rightarrow 0$ as $n\rightarrow 0$.
Our next lemma addresses the remaining two terms in the RHS of~\eqref{eq:P:E}.
\begin{lemma}\label{lem:u:in:L} Let the codeword chosen be $\vec{U}$ (where
$\vec{U}\in \mathcal{T}^n_{\delta}(P_U)$) and let the output on the channel
$W_{Y,Z|X,J}$ be $(\vec{Y},\bZ)$. Then, 

\begin{enumerate}[(a)]
\item there exists $\gamma(\delta)>0$, where $\gamma(\delta)\rightarrow 0$ as $\delta\rightarrow 0$, such that 
 except for an exponentially small probability, $\bU\in \cL_{\gamma(\delta)}(M,\bZ)$. 
\item there exists $f_2(\delta,\epsilon)>0$, where $f_2(\delta,\epsilon)\rightarrow 0$ as $\delta,\epsilon\rightarrow 0$, such that
\end{enumerate}
\begin{IEEEeqnarray}{rCl}
\mathbb{P}\left(\vec{U}_{M,l'}\in \mathcal{L}_{\gamma(\delta)}(M,\vec{Z}), \text{ for some } l'\neq L \right)\leq 2^{-n f_2(\delta,\epsilon)}.
\end{IEEEeqnarray}
\end{lemma}
The proof of this lemma can be found in Appendix~\ref{app:lem:u:in:L}. This
lemma specifies the parameter $\gamma(\delta)$ which appears in the decoder operation in~\eqref{eq:dec:cond}. 
Lemma~\ref{lem:u:in:L} implies that $\bbP(E_{\text{dec},1}|E^c_{\text{enc}}),\bbP(E_{\text{dec},2}|E^c_{\text{enc}}) \rightarrow 0$ as $n\rightarrow 0$.
Hence, we can conclude that $\bbP(E)\rightarrow 0$ as $n\rightarrow
\infty$.

We now get a bound on the expected distortion for $\bx$. Toward this, we first make the following claim.
\begin{claim}\label{claim:tdx}
There exists $r(\delta), f_3(\delta,\epsilon)>0$, where $r(\delta), f_3(\delta,\epsilon)\rightarrow$ as $\delta,\epsilon\rightarrow 0$, such that $\bbP\left((\bx,\btdX)\in \cT^n_{r(\delta)}(P_{X,\wtd{X}})\right)\geq 1-2^{-nf_3(\delta,\epsilon)}$.
\end{claim}
\begin{IEEEproof} 
By Claim~\ref{claim:x:2:u:typ} in App.~\ref{app:lem:u:in:L}, with high probability,
$(\bx,\bJ,\bY,\bZ,\bU)$ is $\delta_4$-typical according to the joint 
distribution $P_{X} T_{\bJ|\bx} W_{Y,Z|X,J} P_{U|Y}$.
As $\vec{\wtd{X}}$ is a deterministic function 
of $(\bU,\bZ)$, it follows by the conditional typicality lemma (see
Lemma~\ref{lem:cond:typ} in Appendix~\ref{app:lem:u:in:L}) that
with probability at least $(1-|\cX||\cJ||\cY||\cZ|
|\cU||\ctdX|2^{-n\delta_4^3})$, the tuple $(\bx,\bJ,\bY,\bZ,\bU,\btdX)$ is
$3\delta_4$-typical, and hence $(\bx,\btdX)$ is
$r(\delta)$-typical, where $r(\delta):=3|\cX| |\ctdX| \delta_4(\delta)$.
This completes the proof.
\end{IEEEproof}
We now show that the \ajblue{maximum distortion} $D^{(n)}$ for the code $\cC$ can be made arbitrarily close to $D$. Let $\bar{E}:=\{(\bx,\btdX)\not \in \cT^n_{r(\delta)}(P_{X,\wtd{X}})\}$. From Claim~\ref{claim:tdx}, we know that $\bbP(\bar{E})\rightarrow 0$ as $n\rightarrow \infty$.
Then, 
\begin{IEEEeqnarray*}{rCl}
\mathbb{E}[d(\vec{x},\vec{\tilde{X}})]   &=&  \mathbb{P}(\bar{E})\mathbb{E}[d(\vec{x},\vec{\tilde{X}})|\bar{E}]+\mathbb{P}(\bar{E}^{c})\mathbb{E}[d(\vec{x},\vec{\tilde{X}})|\bar{E}^c] \\
&\leq& \mathbb{P}(\bar{E})\mathbb{E}[d(\vec{x},\vec{\tilde{X}})|\bar{E}]+\mathbb{E}[d(\vec{x},\vec{\tilde{X}})|\bar{E}^c].
\end{IEEEeqnarray*}
Recall that $d_{\max}<\infty$. In addition, from the typical average lemma~\cite[pg.~26]{elgamal-kim}
we know that $\mathbb{E}[d(\vec{x},\vec{\tilde{X}})|\bar{E}^{c}]\leq D+h(\delta)$, where $h(\delta)>0$ and $h(\delta)\rightarrow 0$ as $\delta\rightarrow 0$. 
Thus, 
\begin{IEEEeqnarray*}{rCl}
\mathbb{E}[d(\vec{x},\vec{\tilde{X}})]   &\leq& \mathbb{P}(\bar{E}) d_{\max}+D+h(\delta)\\
&\stackrel{(a)}{\leq}& D+\epsilon. \yesnumber\label{eq:D:eps}
\end{IEEEeqnarray*}
As $\bbP(\bar{E})\rightarrow 0$ as $n\rightarrow \infty$, we choose a large enough $n$ and a small enough $\delta>0$ to get $(a)$. It follows from~\eqref{eq:D:eps} that the expected distortion for any $\bx\in\cT^n_{\delta_0}(P_X)$ under any randomized $\bJ$ (i.e., any $Q_{\bJ|\bX=\bx}$) is bounded by $D+\epsilon$, and hence, the maximum distortion $D^{(n)}$ can be made arbitrarily close to $D$. We have, thus, shown that for any $\epsilon>0$, the rate $R\leq\max_{Q_{J|X}}(I(U;Y)- I(U;Z))+\epsilon$ is achievable. This completes the proof of achievability.

\section{Proof of Lemma~\ref{lem:u:in:L}}\label{app:lem:u:in:L}
Let us define $\dnot=\delta/2$. Consider the  ``good'' encoder event
$E^c_{\text{enc}}=\{(\bY,\bU)\in\cT^n_{\dtwo}(P_{U|Y}T_{\bY})\}$. We now state and
prove some necessary claims.
\removed{
\begin{claim}
Let $\bX$ be generated i.i.d. $\sim P_X$. Then, with probability at least $(1 - |\cX|e^{-2n\dnot^2})$, $\bX \in 
\cT^n_{\dnot}(P_X)$.
\end{claim}
Let us define this ``good'' event as $A_{X}:=\{\bX \in \cT^n_{\dnot}(P_X)\}$.
}
\begin{claim}\label{lem:j:typ:x}
Given $\bx\in\cT^n_{\delta_0}(P_X)$, and for any $\bj\in\cJ^n$, $(\vec{x},\vec{j})\in \mathcal{T}_{\dnot}^n(P_{X} T_{\vec{j}|\vec{x}})$.
\end{claim}
%
Recall from earlier (cf.~\eqref{eq:D}) our assumption on $\delta_0=\delta_0(n)$.  Here as $\delta\rightarrow 0$, we can make $\delta_0\rightarrow 0$ by making $n\rightarrow \infty$. 

\begin{lemma}[Conditional typicality lemma]\label{lem:cond:typ}
Let $\bs\in\cT^n_{\dnot}(P_S)$ and $\bT$ be generated from $\bs$ using the memoryless distribution $W_{T|S}$. Then,
\begin{IEEEeqnarray}{rCl}
\mathbb{P}\left((\bs,\bT)\in\cT^n_{3\dnot}(P_S W_{T|X})\right)\geq 1-|\cS||\cT| e^{-2 n\delta^3_0}.\label{eq:P:s:t}
\end{IEEEeqnarray}
\end{lemma}
\begin{IEEEproof}
We need to show that
\begin{align*}
&\mathbb{P} \left(\left|T_{\bs,\bT}(s,t)-P_S(s) W_{T|S}(t|s) \right| > 2\dnot\right) 
\end{align*}
is exponentially small for all $s,t$.
We consider two cases. \\
{\it Case I:} $T_{\bs}(s)\leq \dnot$.
As $\bs\in\cT^n_{\dnot}(P_S)$, this implies that 
$ P_S(s)\leq T_{\bs}(s)+\dnot \leq 2\delta_0.  $
Then,~$\forall (s,t)$,
\begin{IEEEeqnarray*}{rCl}
\left|T_{\bs,\bT}(s,t)-P_S(s) W_{T|S}(t|s) \right| &=&\left|T_{\bs}(s)T_{\bT|\bs}(t|s)-P_S(s) W_{T|S}(t|s) \right|\\
&\leq& \max \left( T_{\bs}(s)T_{\bT|\bs}(t|s), P_S(s)W_{T|S}(t|s)\right)  \\
&\stackrel{}{\leq}& 2\dnot \cdot 1  \\
&=&2\dnot.
\end{IEEEeqnarray*}
Thus, for such $s$, $\bbP \left(\left|T_{\bs,\bT}(s,t)-P_S(s) W_{T|S}(t|s) \right| > 2\dnot\right)=0$.\\
{\it Case II:} $T_{\bs}(s)> \dnot$.
Using Chernoff-Hoeffding's  theorem~\cite[Theorem~1]{hoeffding-1963} for each $t\in\cT$, we have
\begin{IEEEeqnarray*}{rCl}
\mathbb{P}(|W_{T|S}(t|s)-T_{\bT|\bs}(t|s)| >\dnot, \text{ for any } t ) \leq |\cT|e^{-2n \delta^3_0 }.
\end{IEEEeqnarray*}
Now, it can be easily checked that 
$|W_{T|S}(t|s)-T_{\bT|\bs}(t|s)| \leq \dnot$ and $|P(s)-T_{\bs}(s)| \leq
\dnot$ together imply 
$$\left|T_{\bs}(s)T_{\bT|\bs}(t|s)-P_S(s) W_{T|S}(t|s) \right| \leq 2\dnot+\dnot^2 \leq 3\dnot.$$
Hence,~\eqref{eq:P:s:t} follows by taking union bound over all $s\in\cS$.
\end{IEEEproof}
Let us denote by $A_{x,J}$, the event that the given $\bx\in\cT^n_{\delta_0}(P_X)$ and the (possibly random) jamming signal $\bJ$ are $\delta_0$-typical w.r.t. $P_X T_{\bJ|\bx}$. Note that it follows from Claim~\ref{lem:j:typ:x} that $\bbP(A_{x,J})=1$ for the specified $\bx\in\cT^n_{\delta_0}(P_X)$.
\begin{claim}\label{claim:A:xjyz}
Conditioned on the event $A_{x,J}$, with probability at least $(1-|\cX||\cJ||\cY||\cZ|e^{-2 \dnot^3 n})$, we have  $(\bx,\bJ,\bY,\bZ)\in\cT^n_{3\delta_0}(P_X T_{\bJ|\bx} W_{Y,Z|X,J})$.
\end{claim}
The proof of this result follows from Lemma~\ref{lem:cond:typ}.
We now consider this ``good'' event $A_{x,J,Y,Z}$, where $A_{x,J,Y,Z}:=\{(\bx,\bJ,\bY,\bZ)
\in \cT_{3\dnot}^n(P_X T_{\bJ|\bX} W_{Y,Z|X,J})\}$.
\begin{claim}\label{claim}
Under the event $A_{x,J,Y,Z}$, $\bY$ is $\done$-typical w.r.t. $P_Y=[P_XT_{\bJ|\bx}W_{Y,Z|X,J}]_Y$, where $\doned:=3|\cX||\cJ||\cZ|\dnotd\rightarrow 0$ as $\delta\rightarrow 0$. That is,
$\|T_{\bY} - P_Y\|_{\infty} \leq 3|\cX||\cJ||\cZ|\dnot$. 
\end{claim}
The proof is straightforward, and hence, omitted. The above claim implies that, except for an exponentially small probability,
the decoder considers the conditional type $T_{\bJ|\bx}$ for decoding.
\begin{claim}\label{claim:y:u:tpy}
Under $E^c_{\text{enc}}$ and $A_{x,J,Y,Z}$, $(\bY,\bU)$ are jointly $\dthree$-typical according to the distribution 
$P_Y P_{U|Y}$, where 
$P_Y=[P_XT_{\bJ|\bx}W_{Y,Z|X,J}]_Y$ and
$\dthreed:=3|\cX||\cY||\cZ|\dnotd + \dtwod \rightarrow 0$ as $\delta \rightarrow 0$.
\end{claim}
\begin{IEEEproof}
Note that
\begin{align*}
\|P_YP_{U|Y} - T_{\bU\bY}\|_{\infty} & \leq \|P_YP_{U|Y} - T_{\bY}P_{U|Y}\|_{\infty}
               + \|T_{\bY}P_{U|Y} - T_{\bU\bY}\|_{\infty}\\
& \leq 3|\cX||\cY||\cZ|\dnot + \dtwo \quad \text{(using $A_{x,J,Y,Z}$ and $E_{\text{enc}}$)} \\
& = \dthree,
\end{align*}
where $\dthree=3|\cX||\cY||\cZ|\dnot + \dtwo$.
\end{IEEEproof}
\begin{claim}\label{claim:unif:dist}
There exists $g(\delta)>0$, where $g(\delta)\rightarrow 0$ as $\delta\rightarrow 0$, such that $\forall \bu\in\cT^n_{\dthree}(P_{U|Y}P_Y|\by)$,
\begin{IEEEeqnarray}{rCl}\label{eq:tdX:dist}
P_{\bU}(\bU =\bu|\bY=\by)\leq 2^{-n(H(U|Y) -g(\delta) )},
\end{IEEEeqnarray}
where $H(U|Y)$ is computed with the distribution $P_{U|Y}P_Y$.
\end{claim}
The proof of this result directly follows from~\cite[Claim~13]{bdp-arxiv2017}.
We now state the following result from~\cite{bdp-it2017} (see also~\cite[Lemma~14]{bdp-arxiv2017}). 
\begin{lemma}[Refined Markov Lemma~\cite{bdp-it2017}~\footnote{  In the refined Markov lemma presented in~\cite{bdp-it2017}~\cite{bdp-arxiv2017}, $K$ depends on $\delta_0$ but does not depend on the block length $n$. Recall that our choice of $\delta_0$ here depends on $n$, which  further results in $K$ also depending on $n$. However, it can be easily verified (see the detailed proof in~\cite{bdp-it2017}) that the first term  in the RHS of~\eqref{eq:K:1} is vanishing in $n$, when $\delta_0=\delta_0(n)\rightarrow 0$ and $\sqrt{n}\delta_0(n)\rightarrow \infty$, as is the case here.}    ]\label{lem:ref:markov:lemma}
Suppose $X\rightarrow Y\rightarrow Z$ is a Markov chain, i.e., $P_{X,Y,Z}=P_{Y}P_{X|Y} P_{Z|Y} $. Let $(\vec{x},\vec{y})\in \mathcal{T}^n_{\dnot}\left(P_{X,Y}\right)$ and $\vec{Z} \sim P_{\vec{Z}}$ be such that
\begin{enumerate}[(a)]
\item $\mathbb{P}\left((\vec{y},\vec{Z})\not\in
\mathcal{T}^n_{\dnot}\left(P_{Y,Z}\right)\right)\leq \epsilon$, where $\epsilon>0$,

\item for every $\vec{z}\in \mathcal{T}^n_{\dnot}\left(P_{Y,Z}|\vec{y}\right)$,
\begin{equation*}\label{eq:condition:markov}
P_{\vec{Z}}(\vec{z})\leq 2^{-n(H(Z|Y)-g(\dnot))},
\end{equation*}
for some $g:\mathbb{R}^{+}\rightarrow\mathbb{R}^{+}$, where $g(\dnot)\rightarrow 0$ as $\delta_0\rightarrow 0$.
\end{enumerate}
Then, there exists $\delta:\mathbb{R}^{+}\rightarrow\mathbb{R}^{+}$, where $\delta (\dnot) 
\rightarrow 0$ as $\dnot \rightarrow 0$, such that
\begin{equation}
\mathbb{P}\left((\vec{x},\vec{y},\vec{Z})\not \in \mathcal{T}^n_{\delta(\dnot)}\left(P_{X,Y,Z}\right)   \right)\leq 2|\cX||\cY||\cZ| e^{-n K} +\epsilon.\label{eq:K:1}
\end{equation}
Here $K>0$ and $K$ does not depend on $P_{X,Y}$, $P_{\vec{Z}}$ or $(\vec{x},\vec{y})$ but does depend on $\dnot$, $g$ and $P_{Z|Y}$. Further, the $\delta$ function does not depend on $(\bx,\by)$, $P_{X,Y}$ or $P_{\bZ}$.
\end{lemma}
The proof of this result follows through careful though minor modifications of the proof in~\cite{bdp-it2017}, and hence, is omitted.
We now use the above lemma to prove the following claim.
\begin{claim}\label{claim:x:2:u:typ}
 There exists $\dfourd>0$, where $\dfourd\rightarrow 0$ as
$\delta\rightarrow 0$, such that except for a small
probability, $(\bx,\bJ,\bZ,\bY,\bU)$
is jointly $\dfour$-typical w.r.t. $P_XT_{\bJ|\bx}W_{Y,Z|X,J}P_{U|Y}$.
\end{claim}
\begin{IEEEproof}
Let us assume that $A_{x,J,Y,Z}$ is true.
Now we use the refined Markov lemma (Lemma~\ref{lem:ref:markov:lemma}) on the Markov chain $(X,J,Z) \rightarrow
Y \rightarrow U$.
Then, by Claims~\ref{claim:A:xjyz}~\ref{claim:y:u:tpy},~\ref{claim:unif:dist}, and Lemma~\ref{lem:covering}, $\bU$ is chosen such that both
conditions (a) and (b) in Lemma~\ref{lem:ref:markov:lemma} are satisfied.
Thus, the claim follows.
\end{IEEEproof}
We define this ``good'' event as $A_{x,J,Y,Z,U}:=\{(\bx,\bJ,\bZ,\bY,\bU)
\in \cT^n_{\dfour}(P_XT_{\bJ|\bx}W_{YZ|XJ}P_{U|Y}\}$. 
\begin{claim}
There exists $\gamma(\delta)>0$, where $\gamma(\delta)\rightarrow 0$ as $\delta\rightarrow 0$, such that except for an exponentially small probability, $\bU\in \cL_{\gamma(\delta)}(M,\bZ)$. 
\end{claim}
\begin{IEEEproof}
Consider the event $A_{x,J,Y,Z,U}$. Under this event, $(\bU,\bZ)$ are
$\gammad$-typical w.r.t. $P_{U,Z}=[P_X T_{\bJ|\bx} W_{Y,Z|X,J}
P_{U|Y}]_{U,Z}$, where $\gammad=|\cX||\cJ| |\cY| \dfour $. 
Thus, the claim follows from Claim~\ref{claim:x:2:u:typ}.
\end{IEEEproof}
%
This completes the proof of the first part of the lemma. The proof of the second part directly follows from the following claim.
\begin{claim}
There exists $f_3(\delta,\epsilon)>0$, where $f_2(\delta,\epsilon)\rightarrow 0$ as $\delta,\epsilon\rightarrow 0$, such that
\begin{IEEEeqnarray}{rCl}
\mathbb{P}\left(\vec{U}_{M,L'}\in \mathcal{L}_{\gamma(\delta)}(M,\vec{Z}), \text{ for some } L'\neq L \right)\leq 2^{-n f_2(\delta,\epsilon)}.
\end{IEEEeqnarray}
\end{claim}
\begin{IEEEproof}
Note that the codewords $\{\bU_{M,L'}\}_{L'\neq L}$ are independently generated, and hence, $\{\bU_{M,L'}\}_{L'\neq L}$ and $\bZ$ are independent. Consider a fixed conditional type $T_{J|X}\in \sQ(T_{\bY})$, and let the resulting distribution $P_{U,Z}=[P_X T_{J|X} W_{Y,Z|X,J} P_{U|Y}]_{U,Z}$. Then,
\begin{align*}
\mathbb{P}\big(\exists l'\neq L:(\vec{U}_{M,l'},\bZ)\in \cT^n_{\gamma(\delta)} (P_{U,Z})) \leq  2^{-n\tdftwo}
\end{align*}
for some $\td{\ftwo} \rightarrow 0$ as $\delta, \epsilon \rightarrow 0$. This follows from the packing lemma~\cite[Lemma~3.1]{elgamal-kim}. 
By taking the union bound over all conditional types $T_{J|X}\in\sQ(T_{\bY})$ (the number of such types is at most polynomial in $n$), we get
\begin{IEEEeqnarray*}{rCl}
\mathbb{P}\big(\exists l'\neq L: (\vec{U}_{M,l'},\bZ)\in \cT^n_{\gamma(\delta)} (P_{U,Z}) \text{  for some } T_{J|X}\in \sQ(T_{\bY})) &\leq& (n+1)^{|\cU||\cZ|} 2^{-n\tdftwo}\\
&\leq & 2^{-n\ftwo}.
\end{IEEEeqnarray*}
\end{IEEEproof}
This completes the proof of the lemma.

%
\bibliographystyle{IEEEtran}

\bibliography{IEEEabrv,References}

\end{document}